\documentclass[prb,twocolumn,showpacs,preprintnumbers,amsmath,amssymb]{revtex4}

\usepackage{epsfig}
\usepackage{graphicx}
\usepackage{dcolumn}
\usepackage{bm}

\begin{document}

\title{Polaron relaxation in a quantum dot due to anharmonic coupling within a mean-field approach}
\author{Tobias Stauber}
\email{tobias.stauber@fisica.uminho.pt}
\author{Mikhail I. Vasilevskiy}
\affiliation{Centro de F\'{\i}sica, Universidade do Minho,
Campus de Gualtar, 4710-057 Braga, Portugal}
\date{\today}

\begin{abstract}
We study the electronic relaxation in a quantum dot within the polaron
approach, by focusing on the {\it reversible} anharmonic decay of
longitudinal optical (LO) phonons forming the polaron into
longitudinal acoustic (LA) phonons. The coherent coupling between the
LO and LA phonons is treated within a mean-field approach. We derive a temperature-dependent inter-level
coupling parameter, related to the Gr\"uneisen parameter and the
thermal expansion coefficient, that characterizes an effective decay
channel for the electronic (or excitonic) states. Within this theory,
we obtain a characteristic anharmonic decay time of 1ns, 2-3 orders of
magnitude longer than previous predictions based on the Fermi's Golden
Rule. We suggest that coherent relaxation due to carrier-carrier interaction is an efficient alternative to the (too slow) polaron decay.
\end{abstract}
\maketitle
{\it Introduction.} The understanding of the relaxation mechanisms in
quantum dots (QDs) is important from both, the fundamental and
technological points of view since QDs are increasingly finding
applications in lasers, single-photon emitters, and possibly quantum
computers where fast carrier dynamics are indispensable. Whereas in
bulk semiconductors, the hot carrier relaxation is mediated via
emission of optical phonons\cite{Zim}, in a QD the discrete electron
and hole levels are separated by energies that generally do not match
any phonon energy. This simple idea gave rise to the "phonon
bottleneck" concept which predicts the inefficiency of hot carrier
relaxation by emission of phonons in QDs\cite{BockBast}. However, this
prediction relies on the assumption that the phonon emission is irreversible,
with a probability described by the Fermi's Golden Rule (FGR).

It has then been realized that multiple scattering processes are
important and that the electron-phonon (e-ph) interaction in QDs must
be treated in a non-perturbative\cite{Hameau} and
non-adiabatic\cite{Fomin} way, leading to the energy spectra described
by polaronic quasi-particle excitations. Polaron spectra of QDs have
been considered theoretically using several approaches
\cite{Stauber,Jacak,Vasilevskiy} and various steady-state observable
properties [such as photoluminescence (PL) and Raman spectra] have been
calculated and compared to experimental data, convincingly demonstrating the
importance of the polaron concept
\cite{Hameau,Fomin,Vasilevskiy,Miranda,Sauvage,Calleja}.

One might then question whether the "phonon bottleneck" in QDs really
exists and, in fact, several experiments point in either direction. On
one hand, an efficient relaxation of optically created electron-hole
pairs (hereafter called excitons) was reported in a number of works studying self-assembled
QDs (SAQDs)\cite{Marcinkevicius,Sun,Muller}, with both PL rise time \cite{Marcinkevicius,Sun} and photoinduced intraband absorption decay time\cite{Muller} below or of the order of 10 ps. Recent
studies\cite{KlimovPRL2005,Hendry} performed on chemically
grown nanocrystals (NCs), where exciton energy level spacings
are larger than in SAQDs, also revealed ultrafast intraband relaxation, 
although the mechanism seems to be different for CdSe and
PbSe NCs. There are some works where the relaxation of a lone carrier
(e.g. electron in an $n$-doped QD) was studied\cite{Sauvage,Zibik,Guyot}. Their results suggest that the
relaxation is slower than for excitons but still fast, with a
characteristic time of several tens of picoseconds for
$n$-doped InAs/GaAs SAQDs\cite{Sauvage,Zibik}.  On the other hand,
there are published experimental results that support the
existence of a phonon bottleneck effect in the relaxation of optically
created excitons, both in self-assembled\cite{Heitz,Urayama} and
nanocrystal\cite{Guyot-Sionnest,Nozik} QDs. For instance, a relaxation
time of 7.7~ns, $\sim $ 15 times the radiation lifetime, was obtained for InAs/GaAs SAQDs in
Ref.~\onlinecite{Heitz}.

The polaron model can explain intraband relaxation of carriers in QDs
only within the context of so-called pseudo-relaxation\cite{stauberB} (i.e. oscillatory dynamics),
since polarons are {\it stationary} states of an electron (exciton)
coupled to optical phonons. Some additional interactions should
therefore be responsible for the true polaron relaxation (i.e. thermalization). Several possible
mechanisms of hot carrier relaxation in QDs have been proposed:

(i) The polaron has a rather short lifetime\cite{Li} because of the
anharmonic effects that lead to a fast decay of confined optical
phonons forming the polaron. This mechanism has been considered for both exciton\cite{Bastard} and lone electron\cite{VerzelenR,Grange} relaxation.

(ii) In SAQDs, the polaron can relax via an Auger-type mechanism assisted
by electrons present in the wetting layer where the energy spectrum is
continuous\cite{Toda,Jahnke}. 

(iii) Acoustic phonons can provide the possibility of transitions
between different (exciton-) polaron states\cite{Vasilevskiy}. If the acoustic
phonon spectrum is (at least partially) continuous and the polaron
spectrum is sufficiently dense, this interaction would drive the 
polaron dynamics towards equilibrium.

An Auger-type mechanism was also proposed\cite{Kharchenko,Zunger} for
the relaxation in chemically grown QDs, according to which the excess
energy is first transferred from the electron to the QD hole through
their Coulomb interaction and the subsequent hole cooling occurs via
emission of acoustic phonons because the hole level spacings
are relatively small and match the continuum of acoustic phonon energies. 
Experimental data support this mechanism, at least for CdSe nanocrystals.\cite{KlimovPRL2005,Hendry,Cooney} 
If one uses the exciton-polaron language, this mechanism is equivalent to (iii).

In this Brief Report, we show that the anharmonicity mechanism (i) is
too slow and therefore not relevant in most of the experimental situations.
It was initially proposed and illustrated for
the simple case of a single longitudinal optical (LO) phonon mode
coupled to two electron states in Ref.~\onlinecite{Li} using the
rotating wave approximation. Later, more realistic calculations of the
polaron spectrum and relaxation rate were performed for an exciton\cite{Bastard} and a lone electron\cite{VerzelenR,Grange} in an InAs/GaAs SAQD. 
In the latter work\cite{Grange}, a good agreement with experimental findings of Ref.~\onlinecite{Zibik} has been achieved by including several anharmonic phonon decay channels known in the literature for various materials\cite{Vallee}.
However, in all these works the LO phonon decay was considered to be irreversible and FGR was used to calculate
the polaron relaxation rate. We shall avoid this approximation, which is known to be a rather crude one for QDs and incoherent with the polaron concept. Instead, we develop a non-perturbative mean-field theory of the QD polaron coupled to acoustic phonons through
{\it reversible} anharmonic processes, resulting in a much slower relaxation. Our main conclusion is that the mechanism (i) is irrelevant in most experimental situations.

{\it The model and approach.} We present a non-perturbative
approach to the non-equilibrium problem of electron relaxation in a
quantum dot. This shall be discussed within a simple (minimal) model
which nevertheless includes all relevant interaction terms. Our QD
contains two non-degenerate electronic levels (separated by an
energy $\epsilon $), coupled to a single confined LO phonon mode. The
LO phonons can {\it reversibly} decay into a couple of longitudinal
acoustic (LA) phonons through an anharmonic process that has been
identified experimentally for some bulk semiconductors\cite{Vallee}.
We shall also include the possibility of further anharmonic
interactions of these (secondary) LA phonons with other vibrational
modes in the QD or its surroundings. The Hamiltonian of this model
can be written as follows:
\begin{align}
H=H_{pol}+H_{LA}+H_{int}+H_{anharm}
\label{H}
\end{align}
where the polaron Hamiltonian is given by
\begin{align}                                            
H_{pol}=\epsilon c_1^\dagger c_1 +\omega_Bb^\dag b+M
(c_1^\dagger c_0  + c_0^\dagger c_1)(b+b^\dagger)\;,
\label{H_polaron}
\end{align} the Hamiltonian of the acoustic phonons reads
\begin{align}
H_{LA}=\omega_A a^\dagger a\;,
\label{H_LA}
\end{align} and the interaction term is
\begin{align}
H_{int}=G(a^\dagger a^\dagger b+b^\dagger aa)\;. 
\label{H_int}
\end{align}
In Eqs. (\ref{H_polaron}-\ref{H_LA}), $\omega_B$ and $\omega_A $ are the energies
of the LO and LA phonons, respectively (we set $\hbar =1$), and $M$ is
the inter-level e-ph coupling constant\cite{Nota1}. In
Eq. (\ref{H_int}), $G$ denotes the characteristic energy of the $LO
\leftrightarrow 2LA$ process related to the experimentally measured anharmonic decay time constant\cite{Vallee,Prabhu}. The last
term in Eq.  (\ref{H}) stands for anharmonic interactions of the
acoustic phonons (excluding those written explicitly as
$H_{int}$). This term is needed for a well-defined ground state (see
below).

In the spirit of a mean-field approach, the interaction term
shall be approximated as
\begin{align}
  H_{int}=&G(\langle a^\dagger a^\dagger\rangle b+b^\dagger \langle
  aa\rangle+a^\dagger a^\dagger \langle b\rangle +\langle b^\dagger
  \rangle aa)\;, \label{H_int2}
\end{align} thus neglecting the coupling
of the fluctuations of the bosonic fields.
By defining shifted bosonic modes, $b\rightarrow b+G\langle a^\dagger a^\dagger\rangle/\omega_B$,
the Hamiltonian (\ref{H}) can be written as
\begin{align}
H=H_{pol}+\widetilde{H}_{LA}+\Delta H-\frac{(G\langle a^\dagger
a^\dagger\rangle)^2}{\omega_B}\;,
\label{H2}
\end{align}
where we have
\begin{align}
\Delta H=-\widetilde M(c_1^\dagger c_0 +c_0^\dagger c_1)
\label{DeltaH}
\end{align}
and
\begin{align}
\widetilde{H}_{LA}=\omega_A a^\dagger a+D(a^\dagger
a^\dagger+aa)\;.
\label{H_LAt}
\end{align}
In Eq. (\ref{DeltaH}), we have introduced a new inter-level coupling
constant,
\begin{align}
\widetilde M=M\frac{2G\langle a^\dagger
a^\dagger\rangle}{\omega_B} \label{Mtil},
\end{align}
and in Eq. (\ref{H_LAt}) an effective coupling constant $D$ including
all the anharmonicity effects involving the LA phonons. This constant takes
into account the effect of the last two terms in Eq. (\ref{H_int2})
as well as $H_{anharm}$ which represents fourth-order processes with the
participation of two LA phonons, i.e., $D=G\langle b\rangle+\widetilde
D$. 

There is a new direct relaxation term $\Delta H$, characterizing the
new relaxation scale $\tau=1/\widetilde M$. To quantify this
scale, we need to know the value of the anomalous amplitude $\langle
a^\dagger a^\dagger\rangle$. The explicit evaluation of the
anomalous amplitude involves the diagonalization of
$\widetilde{H}_{LA}$, which can be made using a Bogoljubov-Valatin
transformation\cite{Mahan},
$
\eta=ga+ha^\dagger,
$
with the normalization of the canonical commutator relations
$g^2-h^2=1$. This gives
\begin{align}
g=\frac{D}{\tilde\omega_A}\frac{1}{h}\;,\;
h=\sqrt{\frac{\omega_A-\tilde\omega_A}{2\tilde\omega_A}}\;,
\label{g_h}
\end{align}
with the renormalized eigenfrequency
\begin{align}
\tilde\omega_A=\sqrt{\omega_A^2-4D^2}.
\label{omega_alphatil}
\end{align}
The occupation number of the acoustic phonons can be expressed in
terms of the expectation value of the new bosons, $\langle
\eta^\dagger \eta\rangle \equiv n_{bos}=(e^{\tilde\omega_A/(k_BT)}-1)^{-1}$, as 
$
n_A\equiv \langle a^\dagger a\rangle=n_{bos}+h^2(1+2n_{bos})\; .
$
The anomalous amplitude entering Eq. (\ref{Mtil}) is given by
\begin{align}
\langle aa\rangle=\langle a^\dagger a^\dagger
\rangle=-gh(1+2n_{bos})\; . \label{a_A}
\end{align}
Thus, all interaction terms in Eq. (\ref{H2}) (except
$H_{pol}$) are now expressed in terms of the effective coupling
constant $D$. In particular, the new inter-level coupling
constant (\ref{Mtil}) is given by
\begin{align}
\widetilde M=-M\frac{2GD} {\tilde\omega_A \omega_B}
(1+2n_{bos})\; . \label{Mtil2}
\end{align}

{\it Self-consistent mean field theory.}
In a self-consistent mean-field theory
involving only electrons and LO and LA phonons, we should put
$D=G\langle b\rangle$. The self-consistency condition is then
defined by the following two equations:
\begin{align}
\langle b\rangle &=-\frac{2G\langle aa\rangle}{\omega_B}\;;\\
\langle aa\rangle &=\frac{-G\langle
b\rangle}{\sqrt{\omega_A^2-4(G\langle b\rangle)^2}}(1+2n_{bos})\;.
\end{align}
This yields the following self-consistent equation for the anomalous
amplitude:
\begin{align}
\omega_A^2\omega_B^2-16G^4\langle aa\rangle^2=4G^4(1+2n_{bos})^2\;.
\end{align}
For $G\to0$, we thus have:
$$\langle
aa\rangle \approx \omega_A\omega_B/(4G^2);\quad
\widetilde M_{MF}\approx M\omega_A/(2G) .
$$ 
This means that the characteristic decay time, $\tau\sim 1/\widetilde M_{MF}$,
tends to zero for vanishing anharmonic
coupling, which certainly is an artifact of the model. The physical
reason for this is the omission of other anharmonic processes
($\widetilde D=0$). It is well known that self-consistent mean-field
theories for boson-fermion systems may lead to such inconsistencies.\cite{Ranninger}

\begin{figure}[tb]
\includegraphics[width=90mm,angle=0]{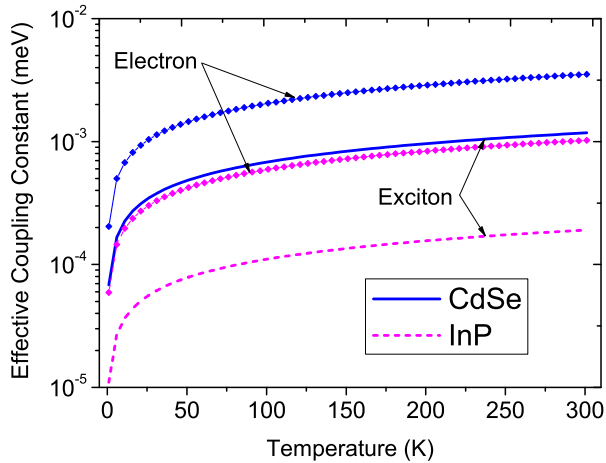}
\caption{(Color online) Effective anharmonic inter-level coupling parameter $\widetilde M$ calculated for excitons and lone electrons in CdSe and InP nanocrystal QDs of 2 nm in radius.}
\label{fig1}
\end{figure}

{\it Effective anharmonic coupling constant.} In order to overcome the
difficulty of the previous section, we have to take into account the
fact that $LO \leftrightarrow 2LA$ is not the only anharmonic process
involving the LA phonons considered in our model. In Eq. (\ref{H_LAt})
we have introduced $D$ as an effective coupling constant representing
all the third- and fourth-order processes with the participation of two LA
phonons. It will now be considered independent of $G$, which is the
coupling constant of one particular anharmonic process. Assuming that
we can ignore other cubic anharmonicity effects (involving just one LA
phonon), we can relate $D$ to the experimentally known Gr\"uneisen
parameter\cite{Born},
$\gamma_{Gr}=-{d\ln \theta }/{d\ln V}$
where $\theta $ is the Debye temperature and $V$ the crystal volume.
In the Gr\"uneisen approximation, the variation of all vibrational
frequencies with the crystal volume is assumed to be the
same\cite{Born}, $d\ln \theta =d\ln \omega $. If we
consider the effective coupling constant as a function of
temperature, $D (T)$, and use Eq. (\ref{omega_alphatil}) for $\tilde\omega_A$, then we can write:
$$
\gamma_{Gr}=-\frac {d\ln \tilde\omega_A}{d\ln V}=-\frac 1 {\alpha
_T}\frac {d\ln \tilde\omega_A}{dT} =-\frac 1 {\alpha
_T} \left ( -\frac 2 {\tilde\omega _A
^2}\frac {dD^2}{dT}\right )
$$
where $\alpha _T $ is the thermal expansion coefficient. Thus, we obtain the following
equation for $D (T)$,
\begin{align}
\frac 2 {\omega _A ^2-4D^2}\frac {dD^2}{dT}=\gamma_{Gr}
\alpha _T\;. 
\label{Gr2}
\end{align}
Its solution, with the condition $D(0)=0$, yields:
\begin{align}
D(T)=\frac {\omega _A} 2 [1- \exp {(-2\gamma_{Gr} \alpha _T
T)}]^{1/2}\;. \label{gammatilde2}
\end{align}
With typical values of $\alpha _T \approx 10^{-5}K^{-1}$ and
$\gamma_{Gr}\approx 1.5$, we obtain $D/\omega _A\approx 1/20$ for $T
\approx 300 K$. Accordingly, $\tilde \omega _A \approx \omega
_A$. Fig. \ref{fig1} shows the temperature dependence of the effective
inter-level coupling parameter $\widetilde M$ following from Eqs.
(\ref{Mtil2}) and (\ref{gammatilde2}), calculated for spherical QDs. We assumed $\omega_A=\omega_B/2$ and used the
anharmonic decay constants from experiment ($G$=0.15 ps$^{-1}$ for CdSe\cite{Prabhu} and 0.04 ps$^{-1}$ for InP\cite{Vallee}) and calculated e-ph coupling parameters\cite{Miranda,Hamma}. $\widetilde M$ thus
turns out to be very small (even for electron in CdSe QD), much smaller then the
typical e-ph interaction energy ($\approx 3$ meV for excitons and $\approx 20$ meV for electrons interacting with spherical-symmetric  phonons). It is larger for lone electron, compared to exciton, because of the stronger coupling to polar phonons.\cite{Nota2}

To conclude, even though the anharmonic decay of LO phonons is faster
in QDs than in bulk semiconductors because of the absence of the
restrictions due to momentum conservation, our results indicate that
the energy scale of this interaction is far below the other relevant
characteristic energies. Consequently, this effective decay
channel is too slow to account for the experimental data. On the contrary, coherent coupling between carriers can produce effective (pseudo-) relaxation of the QD polaron, with $\tau \sim 10$ ps, even if there are no dissipative relaxation channels present.\cite{stauberB}   

{\it Summary.} Assuming the coherent coupling between the optical and acoustic
phonons and using a mean-field approximation, the Hamiltonian
(\ref{H}) describing the interacting polaron can be reduced to $
H=H_{pol}+\Delta H $ with $\Delta H$ given by Eqs. (\ref{DeltaH}),
(\ref{Mtil2}) and (\ref{gammatilde2}). Note that $\Delta H$ is temperature dependent and [apart from
the parameters of the "non-interacting" Hamiltonian
(\ref{H_polaron}, \ref{H_LA})] entirely determined by the thermal
expansion coefficient and the Gr\"uneisen parameter. The term $\Delta
H$ corresponds to a (pseudo-) relaxation time $\tau \sim 1$ ns,
i.e., 2-3 orders of magnitude larger than the previous predictions based on the Fermi's Golden rule. This renders
the anharmonic decay channel virtually irrelevant, at least for
nanocrystal QDs where the polaron is formed by a small number of
confined optical phonon modes most strongly coupled to the electron or exciton.
Instead, the temporal evolution of the photoinduced polarization in a QD can proceed much faster just because of the quantum beats resulting from coherent many-body interactions.\cite{stauberB} In doped SAQD heterostructures, the relaxation of the lone electron or hole occurs due to the coupling to the carriers in the wetting layer,\cite{Jahnke} while in optically excited QDs the same mechanism operates because of the electron-hole interaction\cite{Zunger}. In some cases, this pseudo-relaxation can be taken over by the true (dissipative) one mediated by the continuum of acoustic phonons.\cite{Vasilevskiy} 
In fact, atomistic calculations using the time-domain {\it ab initio}
approach indicate that the relaxation in nanocrystals occurs via e-ph
coupling to acoustic-type vibrations and the phonon bottleneck can
exist only at low energies in very small dots.\cite{Prezhdo} We
finally note that recent experiments\cite{Xu} have shown a bottleneck effect on the hole relaxation in CdSe QDs of 1 - 3 nm in radius (with the gap between the ground and first excited states $\approx $ 30 - 100 meV, similar to SAQDs), which is consistent with the conclusions of this work.

{\it Acknowledgments.}
Support from the FCT (Portugal) through projects PTDC/FIS/64404/2006 and PTDC/FIS/72843/2006 is acknowledged. 



\begin{references}
\bibitem{Nota1}
Note that we omitted the intra-level coupling terms for
simplicity which can be included within a rotating-wave approximation, see Ref. \onlinecite{Stauber}.
\bibitem{Nota2}
Note that this conclusion is in contradiction with the experimental data (see Introduction) and therefore supports the doubts concerning the efficiency of the anharmonic decay mechanism.
\end{references}
\end{document}